\documentclass[preprint,authoryear,10pt]{elsarticle}
\newcommand{\tableItemSpace}{\hspace{0.3cm}}

\usepackage{amssymb}
\usepackage{amsmath} 
\usepackage[nofiglist,notablist]{endfloat} 

\usepackage{ifthen}

\newboolean{BIBUNITRUN}
\setboolean{BIBUNITRUN}{false} 

\ifthenelse{\boolean{BIBUNITRUN}}{ 
    \usepackage{bibunits}
    \defaultbibliography{../../../biblio} 
    \defaultbibliographystyle{model2-names} 
}{  }

\usepackage{lineno}
\usepackage{comment}

\journal{journal}
\begin{document}


\ifthenelse{\boolean{BIBUNITRUN}}{ \begin{bibunit} }{  }

\begin{frontmatter}




\title{Food web framework for size-structured populations}

    \author{Martin Hartvig\corref{Corresponding}\fnref{Lund}}
    \ead{Martin.Pedersen@teorekol.lu.se / martin@hvig.dk}
    \cortext[Corresponding]{Corresponding author.}
    \address[Lund]{Department of Theoretical Ecology, Lund University, Ecology Building, SE-223 62 Lund, Sweden}

    \author{Ken H. Andersen\fnref{Aqua}}
    \ead{kha@aqua.dtu.dk}
    \author{Jan E. Beyer\fnref{Aqua}}
    \ead{jeb@aqua.dtu.dk}
    \address[Aqua]{DTU Aqua, National Institute of Aquatic Resources, Technical University of Denmark (DTU), Charlottenlund Slot, J{\ae}gersborg All\'{e} 1, DK-2920 Charlottenlund, Denmark}

\begin{abstract}
We synthesise traditional unstructured food webs, allometric body size scaling, trait-based modelling, and physiologically structured modelling to provide a novel and ecologically relevant tool for size-structured food webs. The framework allows food web models to include ontogenetic growth and life-history omnivory at the individual level by resolving the population structure of each species as a size-spectrum. Each species is characterised by the trait 'size at maturation', and all model parameters are made species independent through scaling with individual body size and size at maturation. Parameter values are determined from cross-species analysis of fish communities as life-history omnivory is widespread in aquatic systems, but may be reparameterised for other systems. An ensemble of food webs is generated and the resulting communities are analysed at four levels of organisation: community level, species level, trait level, and individual level. The model may be solved analytically by assuming that the community spectrum follows a power law. The analytical solution provides a baseline expectation of the results of complex food web simulations, and agrees well with the predictions of the full model on 1) biomass distribution as a function of individual size, 2) biomass distribution as a function of size at maturation, and 3) relation between predator-prey mass ratio of preferred and eaten food. The full model additionally predicts the diversity distribution as a function of size at maturation.
\end{abstract}


\end{frontmatter}

\section{Introduction}
Food webs are typically modelled using unstructured species populations based on generalised Lotka-Volterra equations. This unstructured formulation ignores individual life-history by assigning a fixed trophic position to all individuals within a species. In aquatic ecosystems this assumption is violated as fish offspring reside at a low trophic level and grow during ontogeny through multiple trophic levels before reaching maturation \citep{bib:Werner_and_Gilliam1984}. Along this journey, from the milligram range and up to several kilogram, fish change diet (as well as enemies) and consequently exhibit life-history omnivory through preying on different trophic levels in different life-stages \citep{bib:Pimm_and_Rice1987}. Thus the assignment of a unique trophic level and role (resource, consumer, predator, etc.) for species in unstructured models is incompatible with systems where ontogenetic growth and life-history omnivory are pronounced. In the cases where trophic level of individuals within a species is positively correlated with body size \citep{bib:Jennings_etal2002a}, individual size may be used as a proxy for trophic level. Models may therefore account for ontogenetic growth and life-history omnivory by resolving the size-structure within each species.

A general framework for large food webs that includes the size-structure for all species must fulfil a set of requirements. It should: 1) be generic in the sense that large species-specific parameter sets are not necessary, 2) be based on mechanistic physiological individual-level processes, where parameters represent measurable biological quantities, 3) resolve food dependent growth of individuals \citep{bib:Werner_and_Gilliam1984}, 4) be practically solvable for species-rich systems over many generations, and 5) comply with empirical data on size-structured communities. In this work we develop a food web framework complying with these requirements by resolving the life-history of individuals within species by a continuous size-spectrum. We parameterise the model for aquatic systems as an example of a size-structured community with widespread life-history omnivory, but the framework may be parameterised for other system types (cf.~Discussion). In fish communities the most prominent empirical patterns, which the model framework should comply with, are that individuals exhibit biphasic growth \citep{bib:Lester_etal2004}, and the Sheldon community spectrum. \cite{bib:Sheldon_etal1972} hypothesised that the community biomass spectrum, from bacteria to whales, as a function of body mass is close to constant. Empiric studies later showed that the biomass for fish indeed is close to constant or slightly declining as a function of body mass \citep{bib:Ursin1982,bib:Boudreau_and_Dickie1992} with the complication that heavily fished systems have a steeper decline in biomass  \citep{bib:Jennings_etal2002a,bib:Daan_etal2005}.

The importance of resolving ontogenetic growth and life-history omnivory has long been realised in fisheries science, where mechanistic individual-level size-structured food web models of fish communities were pioneered \citep{bib:Andersen_and_Ursin1977}. Independently, the physiologically structured population model (PSPM) framework \citep{book:Metz_etal1986,bib:de_Roos_and_Persson2001} has been developed in the field of ecology. While providing the ecological realism needed for a size-structured food web framework these approaches typically rely on large species-dependent parameter sets, which must be reduced for the approaches to be useful as generic frameworks.

Reduction to species-independent parameter sets has been achieved in unstructured models of interacting populations by scaling of physiological and demographic rates with body size \citep{bib:Yodzis_and_Innes1992}. By using body size as a trait this approach has resulted in several simple generic food web models for unstructured populations \citep{bib:Loeuille_and_Loreau2005,bib:Virgo_etal2006,bib:Brose_etal2006b,bib:Lewis_and_Law2007}.

In this work we combine the two approaches into one unified framework: We 1) use a physiological based description of individual life-history, and 2) use a single trait (size at maturation) to characterise each species while using trait and body size scaling to get one condensed species-independent parameter set. All processes are based on descriptions at the level of individuals, and interaction strengths among individuals are dynamic through the prescription of size-dependent food selection. This leads to a realised effective food web structure which depends on the emergent size-spectrum composition of all species. In this manner we synthesise a general framework that in a conceptually simple yet ecologically realistic way can be used to model food webs where the life cycle of individuals in each species is explicitly modelled from birth to reproduction and death.

Our primary objective is the formulation and parametrisation of the food web framework. Food webs generated by unstructured food web models may be analysed at the community level in terms of distributions of biomass across species and trophic levels.  Trait-based size-structured food webs allow a more detailed analysis of the community level as well as enabling analysis on three additional levels of organisation: 1) at the community level, i.e., the distribution of total biomass as a function of body size of individuals regardless of their species identity, and the distribution of biomass and diversity as a function of the trait size at maturation, 2) at the species level, i.e., distribution of biomass as a function of size within a given species, 3) at the trait level, which in the case of a single trait equals the species level, and 4) at the individual level, i.e., distribution of size of food in the stomachs. Due to this added complexity of size-structured food webs, our secondary objective is to illustrate diversity and biomass distributions at different levels of organisation. To this end we generate an ensemble of food webs and analyse them in terms of distributions of average community size-spectra, species size-spectra, trait biomass distributions, and trait diversity distributions. Finally, we develop an analytical solution of the model framework, basically by assuming that the community spectrum follows a power-law (equilibrium size-spectrum theory, EQT). All distributions, except the diversity distribution, may be calculated from EQT, and we demonstrate general accordance between EQT and the results from the full food web simulations. The accordance between EQT and the food web simulations validates the simplifying assumptions behind EQT. EQT provides a ``null-solution'' to the size- and trait-distributions which may be used as a baseline expectation of the results of large size-structured food web simulations.

\begin{figure}[tb]
  \centering
  \includegraphics[width=9.5cm]{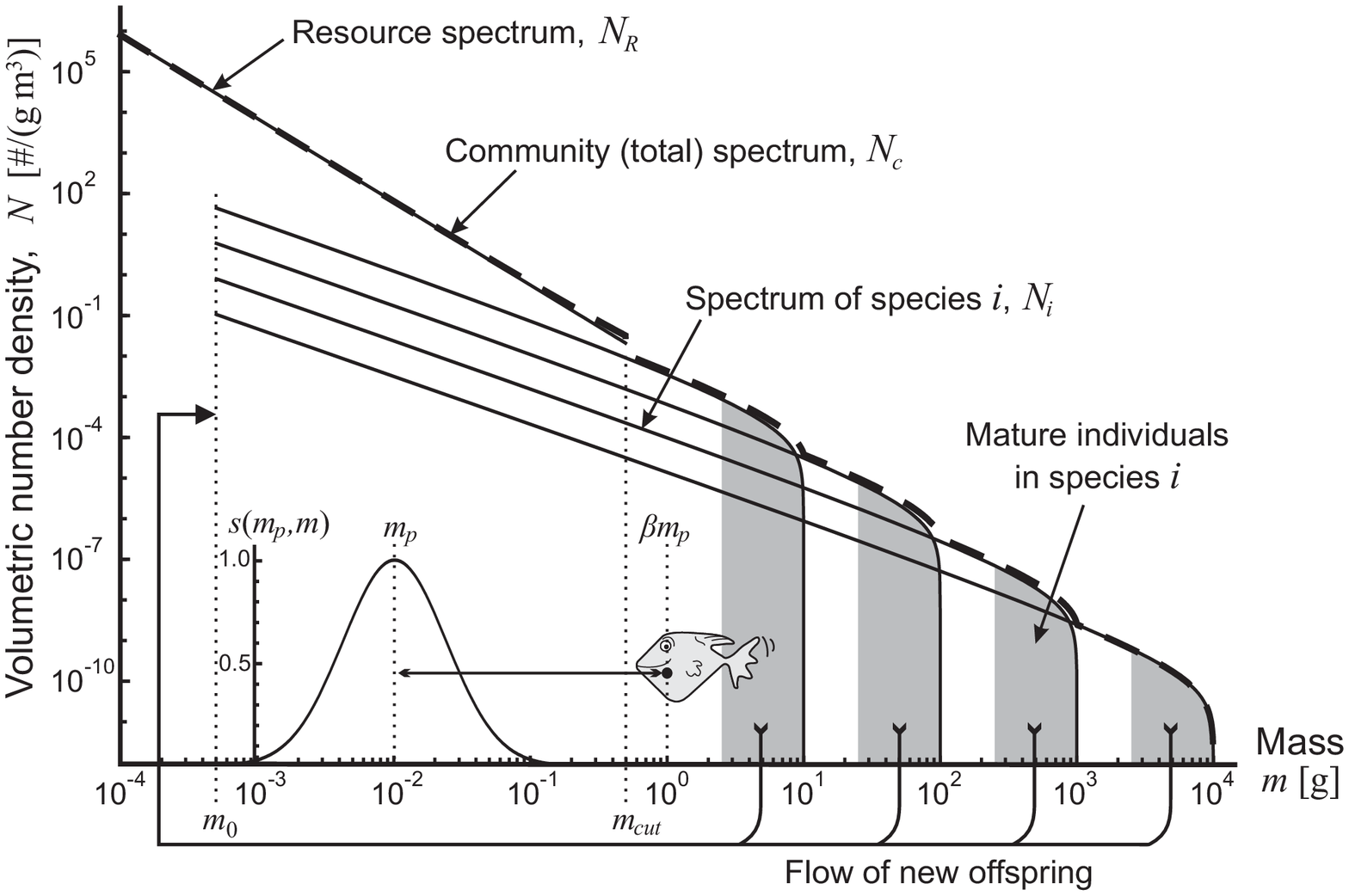}
  \caption{Illustration of the community model: resource spectrum (with cut-off $m_{cut}$) and four species size-spectra ($m_* = 2.5,\ 25,\ 250,$ and $2\,500$ g) having off-spring of size $m_0$. Shaded regions mark the spawning stock from maturation size $m_*$ to maximum asymptotic size $M$. The sum of all spectra gives the community spectrum. The spectra shown are the steady-state solutions from \cite{bib:Andersen_and_Beyer2006}. Inset shows how individuals feed on smaller prey using a feeding kernel with a preferred PPMR of $\beta$.}
  \label{fig:modelsketch}
\end{figure}

\section{Food web model}
The model is based on a description of the processes of food encounter, growth, reproduction, and mortality at the level of an individual with body mass $m$ (Fig.~\ref{fig:modelsketch}). The model is based on two central assumptions: 1) Prey selection is determined at the individual level where individual predators select prey from the rule ``big individuals eat smaller individuals'', and at the species level through introduction of species-specific size-independent coupling strengths \citep{bib:Andersen_and_Ursin1977,bib:Werner_and_Gilliam1984,bib:Emmerson_and_Raffaelli2004}. 2) In addition to species-specific coupling strengths, species identity is characterised by a single trait: size at maturation $m_*$. Interactions among individuals are described by a food encounter process which leads to consumption by predators and mortality on their prey. Food consumption leads to growth in body mass, and when an individual reaches size at maturation $m_*$ it starts allocating energy for reproduction, as well as producing new offspring. Thus the model encapsulates the life-cycle of individuals from birth to maturity and death.

Population dynamics of species $i$ is obtained from individual growth $g_i(m)$ and mortality $\mu_i(m)$ by solving the number conservation equation \citep{bib:McKendrick1926,bib:inBook-vonFoerster1959}:
\begin{equation}
  \frac{\partial N_i}{\partial t} + \frac{\partial}{\partial m}\Big(\, g_i N_i \,\Big) = -\mu_i N_i .
\label{eq:McKendrick-vonFoerster}
\end{equation}
The population structure of species $i$ is described by the size-spectrum $N_i(m,t)$, denoted $N_i(m)$ to ease notation. The size-spectrum represents the volumetric abundance density distribution of individuals such that $N_i(m)\,dm$ is the number of individuals per unit volume in the mass range $[m;\,m+dm]$. Similarly $B_i(m)=mN_i(m)$ denotes the biomass spectrum (biomass density distribution), and $B_i(m)\,dm$ the biomass per unit volume in the range $[m;\,m+dm]$. The sum of all species' size-spectra plus a resource spectrum $N_R(m)$ is the community spectrum (Fig.~\ref{fig:modelsketch}):
\begin{equation}
  N_c(m)=N_R(m) + \sum_i N_i(m).
  \label{eq:community-spectrum}
\end{equation}
The community spectrum represents the entire biotic environment providing individuals with food (from smaller individuals) as well as their predation risk from larger individuals. To include species-specific preferences each species $i$ has its own \emph{experienced} community spectrum:
\begin{equation}
{\cal N}_i(m) = \theta_{i,R} N_R(m) + \sum_j \theta_{i,j} N_j(m) , \label{eq:experienced-community-spectrum}
\end{equation}
\noindent{}where $\theta_{i,j}\in{}[0;\,1]$ is the coupling strength of species $i$ to species $j$. Coupling strengths are independent of body size (cf.~Discussion) since size-dependent food intake is described with a feeding kernel (below).

\subsection{Food consumption}
The consumption of food by an individual depends on the available food from the experienced community spectrum, on the volume searched per time, and on its functional response. The consumed food is assimilated and used to cover respiratory costs. Remaining available energy is used for somatic growth by immature individuals and for a combination of somatic growth and reproduction by mature individuals.

We incorporate the rule of ``big ones eat smaller ones'' by assuming that predators have a preferred predator-prey mass ratio (PPMR). This assumption is inspired by stomach analyses of marine fish \citep{bib:Ursin1973,bib:Ursin1974}, and supported by stable isotope analyses \citep{bib:Jennings_etal2001}. The feeding kernel describing the size preference for prey is prescribed with a normalised log-normal function \citep[Fig.~\ref{fig:modelsketch}, ][]{bib:Ursin1973}:
\begin{equation}
  \label{eq:size_selection}
  s(m_p, m) = \exp\left[ - \left(\ln\left( \frac{\beta m_p}{m} \right)\right)^2 /     (2 \sigma^2)  \right],
\end{equation}
where $m_p$ is prey mass, $m$ predator mass, $\beta$ the preferred PPMR, and $\sigma$ the width of the function. The  food available (mass per volume) for a predator of size $m$ is:
\begin{equation}
  \label{eq:available_food}
  \phi_i(m) = \int m_p {\cal N}_i(m_p) s(m_p,m)\,dm_p.
\end{equation}
Encountered food (mass per time) is the available food multiplied by the volumetric search rate $v(m) = \gamma m^q$, where $q$ is a positive exponent signifying that larger individuals search a larger volume per unit time \citep{bib:Ware1978}. Satiation is described using a feeding level \citep[number between 0 and 1, ][]{bib:Kitchell_and_Stewart1977,bib:Andersen_and_Ursin1977}:
\begin{equation}
  \label{eq:feeding_level}
  f_i(m) = \frac{v(m) \phi_i(m)}{v(m) \phi_i(m) + h m^n},
\end{equation}
\noindent{}where $h m^{n}$ is the maximum food intake. Feeding level times $h m^{n}$ corresponds to a type II functional response.

\subsection{Somatic growth}
Ingested food $f(m)hm^n$ is assimilated with an efficiency $\alpha$ accounting for waste products and specific dynamic action. From the assimilated energy an individual has to pay the metabolic costs of standard metabolism and activity, $km^p$. Thus the energy available for growth and reproduction is:
\begin{equation}
  E_i(m) = \alpha{}f_i(m) h m^{n} - k m^p .   \label{eq:E}
\end{equation}

\noindent{}Of the available energy a fraction $\psi(m,m_*)$ is used for reproduction, and the rest for somatic growth:
\begin{equation}
  g_i(m, m_*) = \left\{ \begin{array}{lc}
   \Big( 1- \psi(m,m_*) \Big) E_i(m) & E_i(m) > 0 \\
   0                                 & \mathrm{otherwise}
\end{array} \right. . \label{eq:growth-function}
\end{equation}
If the intake is insufficient to cover respiratory costs ($E_i(m)<0$) growth is halted. Body size does not shrink when costs cannot be covered, instead starving individuals are exposed to a starvation mortality (see section \ref{sec:mortality}). The maximum asymptotic size $M$ an individual can obtain is reached when all available energy is used for reproduction ($\psi(M,m_*)=1$).

\subsection{Reproduction}
In order to generate growth trajectories with biphasic growth the allocation rule $\psi(m,m_*)$ has to change smoothly from 0 around size at maturation to 1 at the theoretical maximum asymptotic size $M$. The allocation rule $\psi(m,m_*)$ is derived using two requirements: 1) that the size of gonads is proportional to individual mass  \citep{bib:Blueweiss_etal1978}, and 2) that size at maturation is proportional to asymptotic size \citep{bib:Beverton1992,bib:Froese_and_Binohlan2000,bib:He_and_Stewart2001}. To obtain an analytical solution as to how individuals allocate available energy to growth and reproduction we assume that the allocation rule is based on a constant feeding level $\overline{f}$. Requiring allocation to reproduction to be proportional to individual mass, $\psi(m,m_*) \overline{E}(m) =k_r m$, gives $\psi(m,m_*)=k_r m/\overline{E}(m)$, where $\overline{E}(m)=\alpha{} \overline{f} h m^{n} - k m^p$ denotes the available energy when feeding level is constant. The factor $k_r$ is found by the second requirement through $\psi(M,m_*)=1$: $k_r = \overline{E}(M)/M$ where $M=m_*/\eta_*$. The allocation can thus be described as:
\begin{equation}
  \psi(m,m_*) =  \left[ 1+\left( \frac{m}{m_*}\right)^{-u}\right]^{-1}     \frac{\overline{E}(m_*/\eta_*)}{\overline{E}(m)} \frac{m}{m_*/\eta_*},    \label{eq:psi-function}
\end{equation}
\noindent{}where the term in the square brackets is a smooth step function switching from zero to one around the size at maturation ($u$ determines transition width).

The exponents of maximum consumption and standard metabolism are close to equal (cf.~\ref{app:parameter_estimation} and Discussion). In the limit of $n=p$ the available energy for growth and reproduction becomes $\overline{E}(m)=\hbar m^n$ where $\hbar = \alpha \overline{f} h-k$. This gives: $\psi(m,m_*) = [1+( m/m_*)^{-u}]^{-1} (\eta_*m/m_*)^{1-n}$, meaning that the juvenile growth pattern is $g=\hbar m^n$ whereas adults grow according to $g=\hbar m^n - \hbar (m_*/\eta_*)^{n-1}m$. Thus the growth model is a biphasic growth model where adults follow von Bertalanffy growth curves as advocated by \cite{bib:Lester_etal2004}.

The total flux of offspring is found by integrating the energy allocated to reproduction $\psi(m,m_*)E_i(m)$ over all individual sizes:
\begin{equation}
  R_i = \frac{\epsilon}{2m_0}\int N_i(m) \psi(m,m_*)E_i(m) \, dm,   \label{eq:reproduction}
\end{equation}
where $m_0$ is the egg size, $\epsilon$ the efficiency of offspring production (\ref{app:epsilon_deriv}), and 1/2 takes into account that only females spawn (assuming equal sex distribution). Reproduction determines the lower boundary condition of \eqref{eq:McKendrick-vonFoerster} for the size-spectrum of the species:
\begin{equation}
  \label{eq:BC}
g_i(m_0, m_*) N_i(m_0) = R_i.
\end{equation}

\subsection{Mortality} \label{sec:mortality}
The mortality rate $\mu(m)$ of an individual has three sources: predation mortality $\mu_p(m)$, starvation mortality $\mu_s(m)$, and a small constant background mortality $\mu_b(m_*)$. The background mortality is needed to ensure that the largest individuals in the community also experience mortality as they are not predated upon by any individuals from the community spectrum.

Predation mortality is calculated such that all that is eaten translates into predation mortalities on the ingested prey individuals (\ref{app:pred-mortality}):
\begin{equation}
  \label{eq:mu}
  \mu_{p,i}(m_p) = \sum_j \int s(m_p,m) (1-f_j(m)) v(m) \theta_{j,i} N_j(m)\, dm.
\end{equation}

When food supply does not cover metabolic requirements $k m^p$ starvation mortality kicks in. Starvation mortality is proportional to the energy deficiency $k m^p - \alpha f(m) h m^n$, and inversely proportional to lipid reserves, which are assumed proportional to body mass:
\begin{equation}
  \label{eq:mu_s}
  \mu_s(m) = \left\{ \begin{array}{lc}
             0                      & E_i(m) > 0 \\
             \frac{-E_i(m)}{\xi{}m} & \mathrm{otherwise}
\end{array} \right. .
\end{equation}

Mortality from other sources than predation and starvation is assumed constant within a species and inversely proportional to generation time \citep{book:Peters1983}:
\begin{equation}
  \label{eq:mu_b}
  \mu_b = \mu_0 m_*^{n-1}.
\end{equation}

\subsection{Resource spectrum}
The resource spectrum $N_R(m)$ represents food items which are needed for the smallest individuals (smaller than $\beta m_0$). The dynamics of each size group in the resource spectrum is described using semi-chemostatic growth:
\begin{equation}
  \label{eq:nb}
  \frac{\partial N_R(m,t)}{\partial t} = r_0m^{p-1} \Big[ \kappa m^{-\lambda} - N_R(m,t) \Big] - \mu_p(m) N_R(m,t),
\end{equation}
where $r_0m^{p-1}$ is the population regeneration rate \citep{bib:Fenchel1974,bib:Savage_etal2004} and $ \kappa m^{-\lambda}$ the carrying capacity. We prefer semi-chemostatic to logistic growth since planktonic resources rebuild from depletion locally due to both population growth and invasions.

\begin{table}[tb]
\centering
\begin{small}
  \caption{Default parameter values for a temperature of $10^\circ$C (\ref{app:parameter_estimation}).}
  \label{tab:parameters}
  \begin{tabular}{@{\hspace{0.1cm}}l@{\hspace{0.15cm}}l@{\hspace{0.2cm}}l@{\hspace{0.2cm}}l@{\hspace{0.2cm}}l@{\hspace{0.1cm}}}
    \hline
    \hline
    & Symbol & Value & Units & Parameter \\
    \hline
    \multicolumn{5}{@{\hspace{0.1cm}}l}{ Individual growth} \\
    & \tableItemSpace$f_0$ & 0.6 & - & Initial feeding level\\
    & \tableItemSpace$\alpha$ & 0.6 & - & Assimilation efficiency\\
    & \tableItemSpace$h$ & 85 & $\mathrm{g}^{1-n}$/year & Maximum food intake \\
    & \tableItemSpace$n$ & 0.75 & - & Exponent for max. food intake \\
    & \tableItemSpace$k$ & 10 & g$^{1-p}/$year & Std. metabolism and activity \\
    & \tableItemSpace$p$ & 0.75 & - & Exponent of std. metabolism \\
    & \tableItemSpace$\beta$ & 100 & - & Preferred PPMR \\
    & \tableItemSpace$\sigma$ & 1 & - & Width of feeding kernel \\
    & \tableItemSpace$q$ & 0.8 & - & Exponent for search volume \\
    \multicolumn{5}{@{\hspace{0.1cm}}l}{ Reproduction} \\
    & \tableItemSpace$m_0$ & 0.5 & mg & Offspring mass \\
    & \tableItemSpace$\eta_*$ & 0.25 & - & $m_*$ rel. to asymptotic mass $M$  \\
    & \tableItemSpace$\epsilon$ & 0.1 & - & Efficiency of offspring production \\
    & \tableItemSpace$u$ & 10 & - & Width of maturation transition\\
    \multicolumn{5}{@{\hspace{0.1cm}}l}{ Mortality} \\
    & \tableItemSpace$\xi$ & 0.1 & - & Fraction of energy reserves \\
    & \tableItemSpace$\mu_0$ & 0.84 & g$^{1-n}/$year & Background mortality \\
    \multicolumn{5}{@{\hspace{0.1cm}}l}{ Resource spectrum} \\
    & \tableItemSpace$\kappa$ & $5\cdot{}10^{-3}$ & g$^{\lambda-1}/$m$^3$ & Magnitude of resource spectrum \\
    & \tableItemSpace$\lambda$ & $2-n+q$ & - & Slope of resource spectrum \\
    & \tableItemSpace$r_0$ & 4 & g$^{1-p}$/year & Regeneration rate of resources\\
    & \tableItemSpace$m_{cut}$ & 0.5 & g & Upper limit of resource spectrum \\

    \hline
    \hline
  \end{tabular}
\end{small}
\end{table}

\subsection{Derivation of parameters}
Each species is characterised by a single trait, size at maturation $m_*$, and a species-independent parameter set is achieved through scaling with body size $m$ and $m_*$. The model is parameterised for marine ecosystems using cross-species analyses of fish communities (\ref{app:parameter_estimation} and Table \ref{tab:parameters}).

The constant $\gamma$ for the volumetric search rate is difficult to assess (\ref{app:parameter_estimation}). However, since the feeding level $f(m)$ of small individuals is determined solely by the amount of encountered food from the resource spectrum, we may use \emph{initial feeding level} $f_0$ as a  physiological measure of food encounter; $f_0$ is defined as the feeding level resulting from a resource spectrum at carrying capacity. The initial feeding level is used as a control parameter for food availability (enrichment), through which the value of $\gamma$ can be calculated (\ref{app:gamma}):
\begin{equation}
\gamma_i(f_0) = \frac{f_0 h \beta^{2-\lambda}}{(1-f_0)\sqrt{2\pi}\theta_{i,R}\kappa\sigma},
\end{equation}
\noindent{}where it is noted that $\gamma$ will be species dependent if species have different coupling strengths to the resource.

A \emph{critical feeding level} $f_c$ can be formulated as the feeding level where all assimilated food is used for metabolic costs  (using values from Table \ref{tab:parameters}):
\begin{equation}
  f_c = \frac{k}{\alpha h}m^{p-n} =  \frac{k}{\alpha h} \approx 0.2 .
  \label{eq:fc}
\end{equation}
Individuals can only grow and reproduce if $f>f_c$. Assuming that individuals experience an average feeding level $\overline{f}$, the growth \eqref{eq:growth-function} of juveniles is $g=\hbar m^n$  (for $n = p$). The parameter $\hbar=\alpha h \overline{f} - k$ can be estimated through the relation between observed von Bertalanffy growth rate and asymptotic size yielding $\hbar\approx10\,\mathrm{g}^{0.25}$/year \citep{bib:Andersen_etalMP2008}. This allows an estimation of the expected average feeding level of individuals in the field (Table \ref{tab:parameters}):
\begin{align}
\overline{f} = \frac{\hbar + k}{\alpha h} \approx 0.4 ,
\end{align}
i.e.~around twice the critical feeding level. As the initial feeding level $f_0$ is calculated from a resource spectrum at carrying capacity, the realised feeding level in the model will be smaller than $f_0$.  A value of $f_0=0.6$ was seen to give realised feeding levels around 0.4.

\section{Methods} \label{sec:methods}
Stable food webs are constructed using the full dynamic food web model with random coupling strengths $\theta_{i,j}$. For each run, 30 species are assigned with $m_*$ evenly distributed on a logarithmic size axis ($m_* \in [0.25\,\mathrm{g};\, 20\,\mathrm{kg}]$), random $\theta_{i,j}$ matrices (mean 0.5), and a common $\theta_{i,R}=0.5$ coupling to the resource spectrum. Numerical integration is performed by standard finite difference techniques (\ref{app:numerical_method}). Food webs are simulated in 10 consecutive intervals covering 300 years each, where species with a biomass less than $10^{-30}\mathrm{g/m^3}$ are eliminated after each interval. To eliminate food webs that still have not reached the final state each community is integrated for additional 500 years and discarded if any species has an absolute population growth rate larger than 1 logarithmic decade per 100 years. To ensure that each food web in the final ensemble spans multiple trophic levels we only retain food webs where at least one species has $m_*$ larger than 2.5 kg. For statistics we use the mean of the last 250 years of the simulation with time steps saved in 0.1 year increments. In this manner 204 food webs having a total number of 1016 species were collected. Each web contained between 2 and 9 species with a mean of 4.98 species.

We analyse the generated food webs in terms of distributions of average community size-spectrum, species size-spectra, trait biomass distributions, and trait diversity distributions.  Additionally we demonstrate the importance of distinguishing between what an individual prefers to eat and what is actually ingested (i.e.~found in its stomach) by showing how emerging PPMRs vary with food availability and differ from preferred PPMRs.

An approximate steady-state solution to the food web model which neglects the dynamics of reproduction can be found using two assumptions: 1) all species consume food and experience mortality from a scaling community size-spectrum $N_c = \kappa_c m^{-\lambda}$, and 2) constant feeding level $\overline{f}$, which implies equal species coupling strengths $\theta_{i,j}=\overline{\theta}$. Whereas the food webs in the full model are based on a discrete set of $m_*$, the analytical solution considers $m_*$ as a continuous distribution.  The procedure for deriving the analytical solution is similar to the derivation of equilibrium size-spectrum theory \citep{bib:Andersen_and_Beyer2006}, but the results are slightly different as standard metabolism is taken explicitly into account here. The food encountered by an individual is found using assumption 1): $v(m)\phi(m) = \gamma m^q \int N_c s(m_p,m)m_p\,dm_p \propto m^{2-\lambda+q}$. The feeding level is calculated from \eqref{eq:feeding_level}, and the requirement that it is constant (assumption 2) leads to a constraint on the exponent of the community spectrum: $\lambda = 2+q-n$. Feeding with a constant feeding level generates a predation mortality of $\mu_p = \alpha_p m^{n-1}$ (\ref{app:pred-mortality}). The size-spectrum of juvenile individuals is found as the steady state solution of \eqref{eq:McKendrick-vonFoerster} using the above predation mortality and $g = \hbar m^n$ (cf. \eqref{eq:app:N_solution}): $N(m,m_*) = \kappa(m_*) m^{-n-a}$, where $a = \alpha_p/\hbar$ is the physiological level of predation \citep{bib:Beyer1989,bib:Andersen_and_Beyer2006}, which can be calculated as $a \approx \overline{f}/(\overline{f}-f_c) \beta^{2n-q-1} / \alpha = 0.86$ (\ref{app:avail_food_and_a}). The constant $\kappa(m_*)$ is found from the requirement that the sum of all species spectra should equal the community spectrum. Assuming a continuum of species the requirement can be written as $\int N(m, m_*)\, \mathrm{d}m_* = N_c(m)$ which leads to $\kappa(m_*) \propto m_*^{2n-q-3+a}$ (Fig.~\ref{fig:modelsketch}). This approximate solution of the model will be referred to as equilibrium size-spectrum theory (EQT), and it will be compared to the output of the complete dynamic food web model.

In dynamic models, as in nature, the lifetime reproductive success (fitness) has to be $R_0 = 1$ for all coexisting species. Since EQT does not consider the boundary condition \eqref{eq:BC} life-time reproductive success becomes a function of size at maturation: $R_0 \propto m_*^{1-a}$ \citep{bib:Andersen_etalMP2008}. One solution to making $R_0$ independent of $m_*$ is to set $a=1$, but that breaks the above employed mass balance between growth and mortality used to calculate $a$. Due to the $R_0 \neq 1$ inconsistency in EQT we have a specific focus on the realised values of $a$ when comparing food web simulations with EQT predictions. To examine how the regulation of $R_0$ occurs in the full food web model $R_0$ is split into two factors: 1) the probability of surviving to become adult, and 2) lifetime reproduction per adult (\ref{app:EQT}):
\begin{equation}
p_{m_0\rightarrow{}m} = \frac{N(m)g(m, m_*)}{N(m_0)g(m_0, m_*)}, \label{eq:survival}
\end{equation}
\begin{equation}
R_\mathrm{adult}(m_*)=\int_{m_*}^M p_{m_*\rightarrow{}m} \frac{\psi(m,m_*)E(m)}{g(m, m_*)} dm. \label{eq:R_life}
\end{equation}

Survival probabilities and reproductive outputs in the food web simulations are compared with EQT predictions, which are calculated by inserting the EQT size-spectra into \eqref{eq:survival} and \eqref{eq:R_life}. Juvenile growth is $g\propto{}m^n$, which gives $p_{m_0\rightarrow{}m_*} \propto m_*^{-n-a}m_*^{n}=m_*^{-a}$ and $R_\mathrm{adult}\propto{} m_*$.

\begin{figure}[t]
  \centering
  \includegraphics[width=12cm]{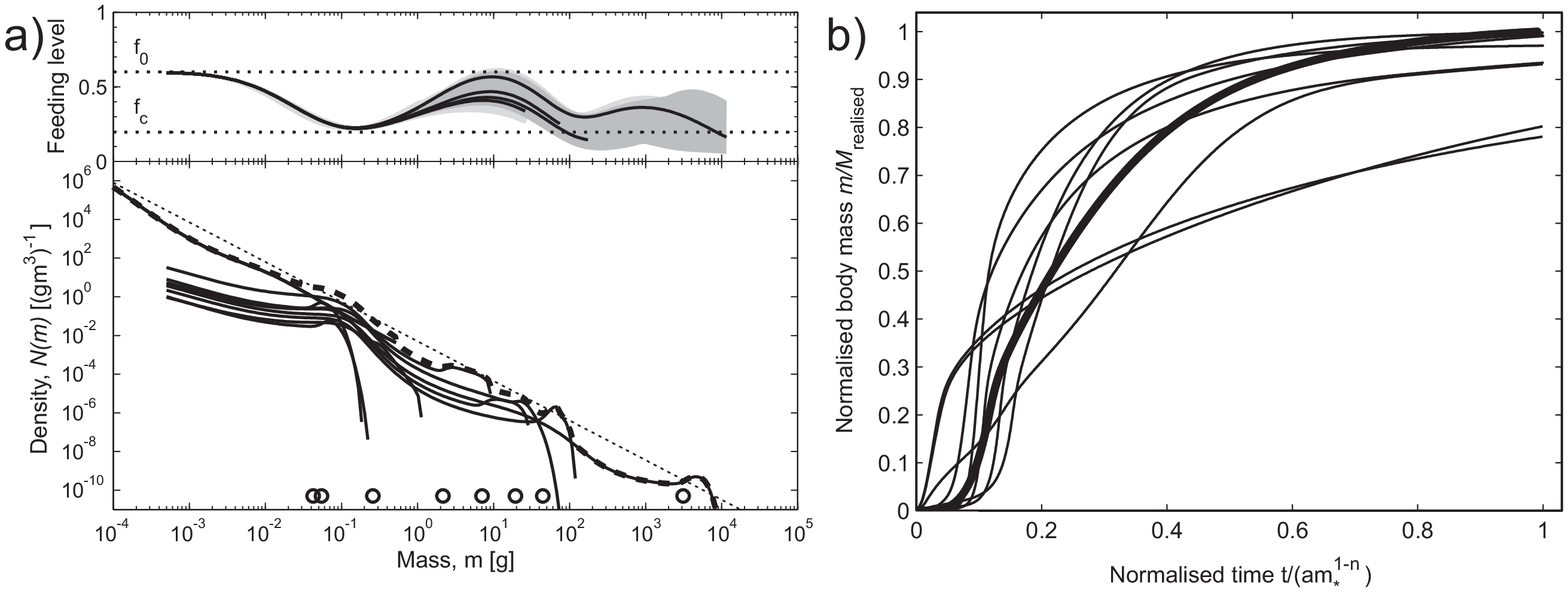}
  \caption{Example of an eight species cyclic state. a) Top: Feeding levels of the species along with the min/max (light grey) and the 25\%/75\% (dark grey) percentile values of the time-series. Dashed lines indicate initial $f_0$ and critical $f_c$ feeding level. Bottom: The time-average of the resource and species spectra along with the community spectra (thick dashed). The idealised community spectrum $\kappa_c m^{-\lambda}$ (thin dashed) and the species maturation sizes $m_*$ (circles). b) Time averaged growth curves for the species (thin lines) along with the biphasic growth curve \eqref{eq:growth-function} for a fixed feeding level that equals 75\% of the time and size averaged feeding level experienced by the species (thick line). Growth curves are normalised with realised asymptotic size ($y$-axis) and generation time ($x$-axis) to enable comparison.}
  \label{fig:multi-species-state}
\end{figure}

\section{Model predictions}

\subsection{Growth trajectories}
In unstructured models fluctuations are manifested as oscillations in the biomass of species, whereas the oscillations in structured models stem from oscillations in the size-spectrum composition. Such oscillations give rise to fluctuating feeding levels as individuals encounter different levels of food in different life-stages (Fig.~\ref{fig:multi-species-state}.a).  Variations in feeding levels between species and as a function of individual size lead to different emergent growth trajectories (Fig.~\ref{fig:multi-species-state}.b).  The growth trajectories roughly follow the biphasic growth curve that is obtained if the feeding level is assumed to be constant.

\begin{figure}[tb]
  \centering
  \includegraphics[width=7cm]{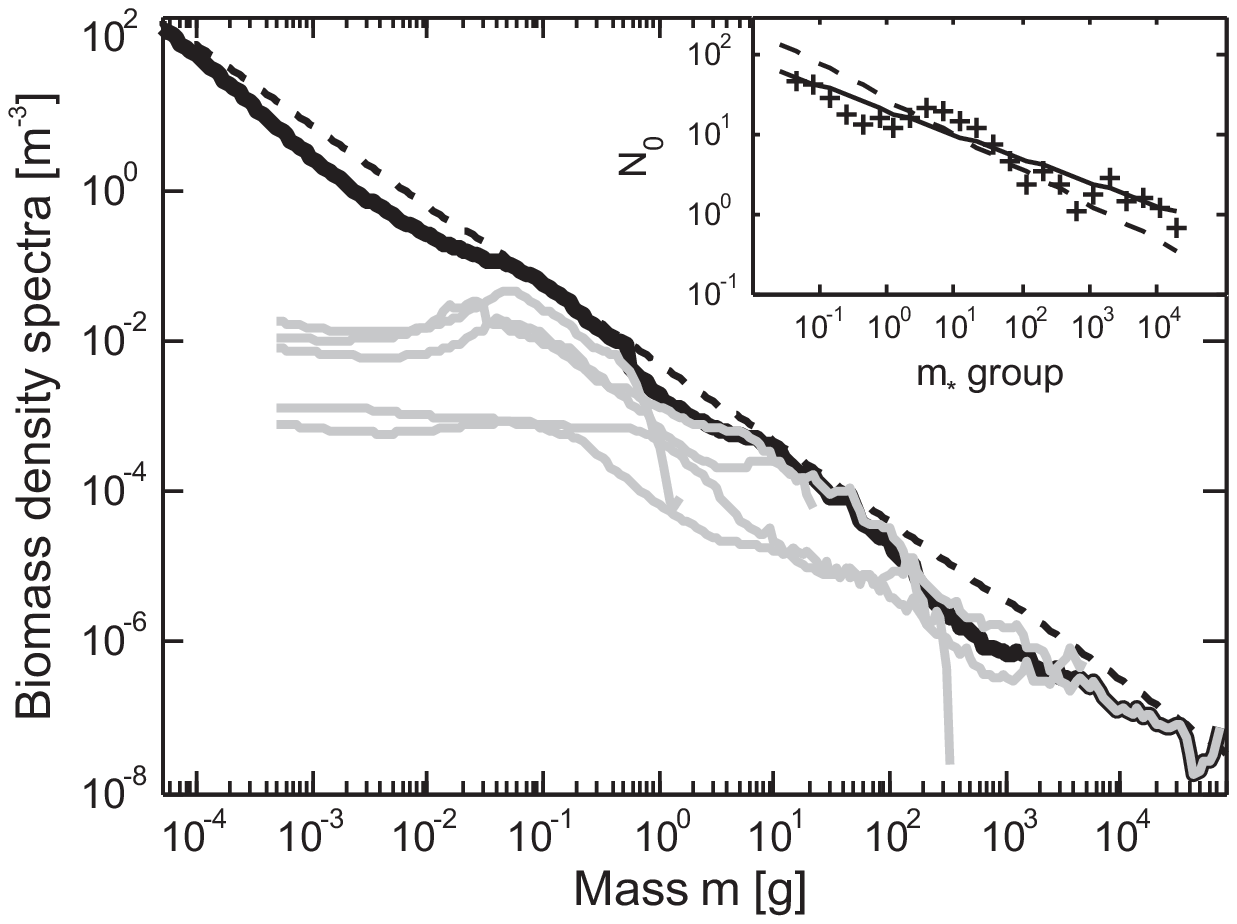}
  \caption{Mean species biomass spectra (grey lines) when species are divided into 5 logarithmic evenly distributed $m_*$ groups. Also shown is the total mean community biomass spectrum (thick line), and the EQT community biomass spectrum $\kappa_c{}m^{1-\lambda}$ (dashed). Inset shows how offspring abundance $(N_0)$ scales with $m_*$ (data pooled in 25 log groups). Expected EQT scaling of $N_0$ is shown for $a=0.86$ (dashed) and $a=1$ (solid).
  }
  \label{fig:spectra}
\end{figure}

\begin{figure}[tb]
  \centering
  \includegraphics[width=12cm]{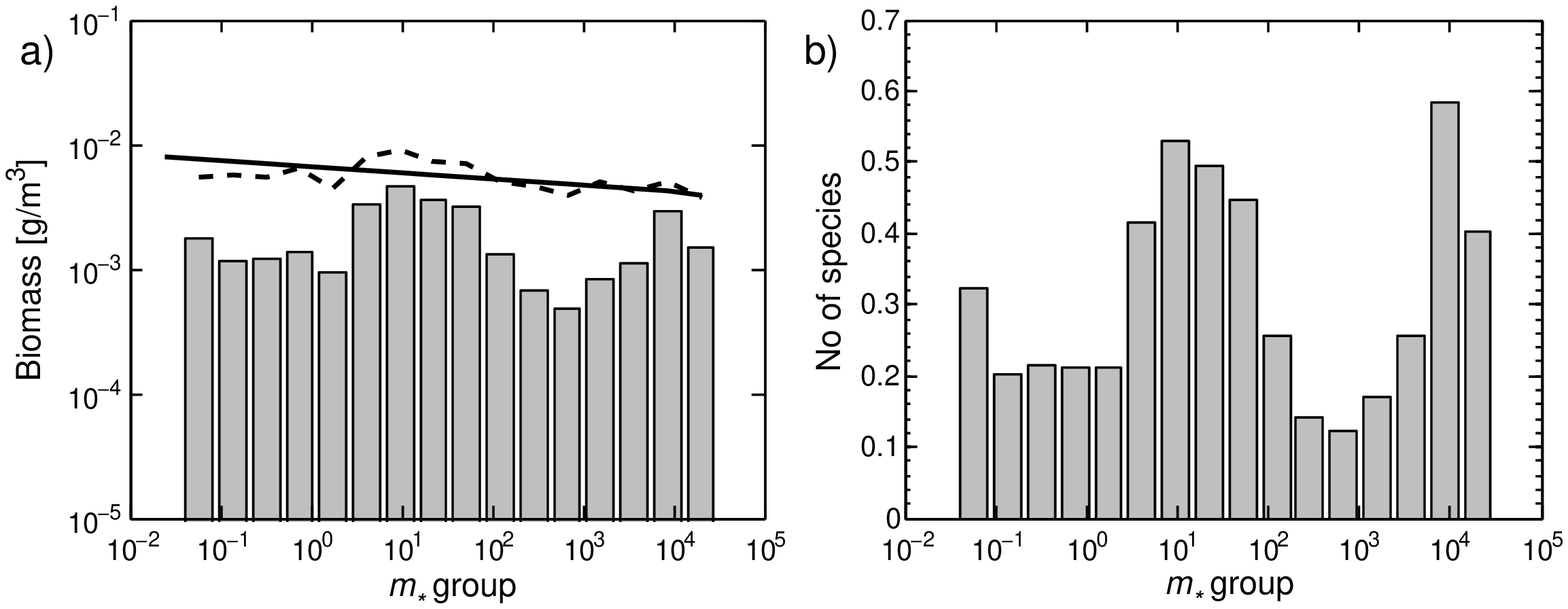}
  \caption{a) Distribution of biomass in different $m_*$ groups. The expected distribution $B(m_*) \propto m_*^{n-q}$ is illustrated with the solid line. Dashed line shows biomass per species (bar values divided with bar values in b)). b) Mean no of species as a function of $m_*$. Species are pooled into 16 logarithmic groups.}
  \label{fig:biomassdistribution}
\end{figure}

\subsection{Biomass structure}
By pooling species from each food web into logarithmic evenly distributed $m_*$ groups, and summing the size-spectra in each group, a size-spectrum is obtained for each $m_*$ group. Next, the logarithmic average of $m_*$ groups across all food webs is performed to produce the average size-spectra of a $m_*$ group (Fig.~\ref{fig:spectra}). Average community biomass spectrum $N_c(m)m$ follows the EQT prediction of a slope of $1+q-n=1.05$, meaning that the biomass in logarithmically evenly sized size-groups, $\int_m^{cm} N_c(m) m dm$, is a slightly declining function of body mass. The community spectrum oscillates around the EQT prediction due to a trophic cascade initiated by a superabundance of the largest predators which do not experience any predation mortality \citep{bib:Andersen_and_Pedersen2010}. The peaks of the oscillating pattern are roughly spaced by the preferred PPMR. Biomass density within species is constant until individuals reach the end of the resource spectrum, and larger individuals, $\gtrsim 0.1$\,g, have a biomass spectrum slope flatter than that of the community spectrum (Fig.~\ref{fig:spectra}). Thus, in contrast to EQT, the dynamic model produces species size-spectra that cannot be described as power laws. The number of small individuals is inversely related to size at maturation. The scaling of offspring abundance can be calculated using EQT as $N_0\propto{}\int_{m_*}^{cm_*}\kappa(m'_*)dm'_*\propto{}m_*^{2n-q-2+a}$, which fits the simulated results well for $a=1$ (Fig.~\ref{fig:spectra}, inset).

The distribution of species biomass as a function $m_*$ can be calculated from EQT as:
\begin{equation}
B(m_*) = \int_{m_*}^{cm_*} \int_{m_0}^M N(m,m'_*) m dm \, dm'_* \propto m_*^{n-q}. \label{eq:ExtendedSheldon}
\end{equation}
As $n$ and $q$ are almost equal the biomass distribution $B(m_*)$ as a function of $m_*$ is almost constant. This result is also borne out by the dynamical simulations (Fig.~\ref{fig:biomassdistribution}.a) with some variation due to uneven species distribution along the $m_*$ axis: peaks occur in species diversity separated by the preferred PPMR $\beta$ (Fig.~\ref{fig:biomassdistribution}.b).

\begin{figure}[tb]
  \centering
  \includegraphics[width=12cm]{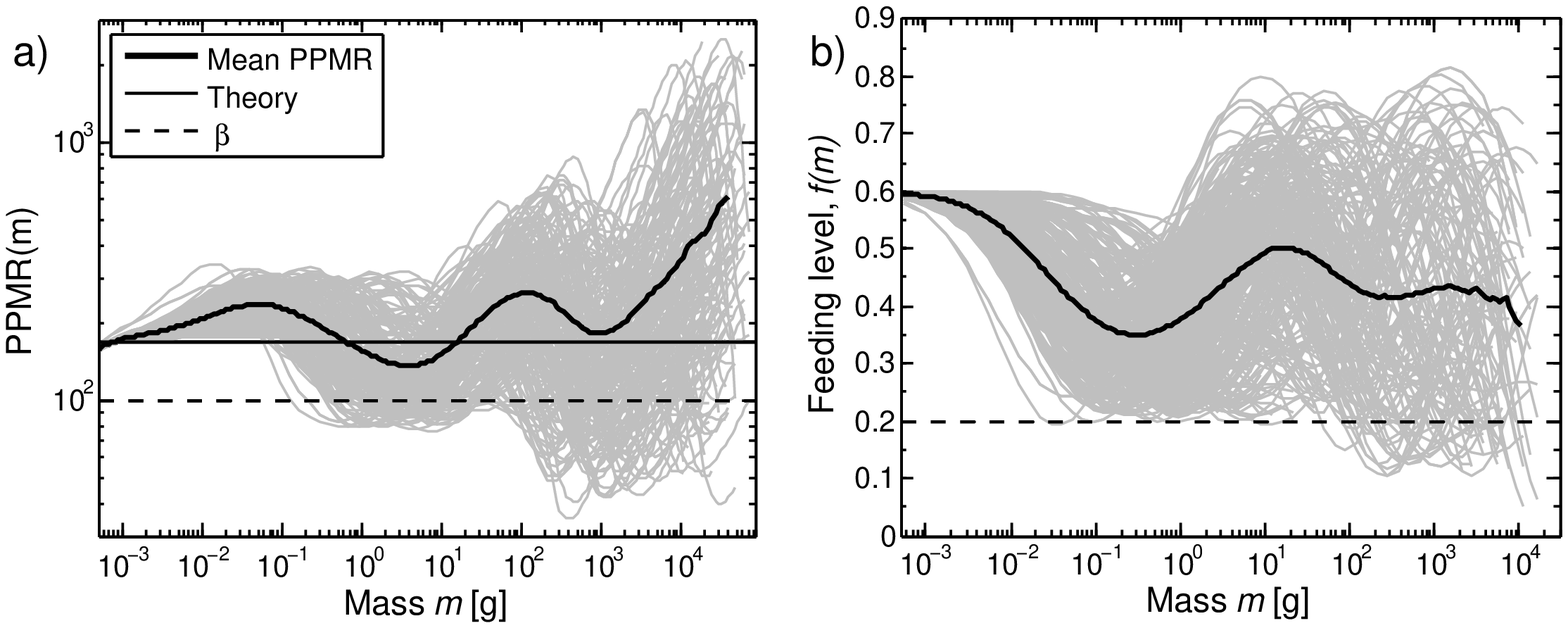}
  \caption{a) Realised PPMR from each food web (grey), mean realised PPMR across all simulations (thick black), realised PPMR prediction from equilibrium theory (thin black), and preferred PPMR $\beta$ (thin dashed). b) Feeding level from each species (grey), and mean feeding level (thick black).}
  \label{fig:PPMR_feedinglevel}
\end{figure}

\subsection{PPMR and feeding level}
The realised mean PPMR can be derived when prey concentrations are known: ${\cal N}(m_p)s(m_p,m)$ is the prey size distribution encountered by a $m$ sized predator. Mean prey size encountered by a $m$ sized predator is $\frac{\int_0^\infty m_p {\cal N}(m_p)s(m_p,m) dm_p}{\int_0^\infty{\cal N}(m_p)s(m_p,m)dm_p}$. The realised mean PPMR is calculated as the predator size $m$ divided by the mean prey size:

\begin{equation}
PPMR(m) = \frac{m \int_0^\infty {\cal N}(m_p)s(m_p,m) dm_p}{\int_0^\infty m_p {\cal N}(m_p)s(m_p,m) dm_p} . \label{eq:PPMR}
\end{equation}

Realised mean PPMR is always larger than the preferred PPMR $\beta$, due to higher abundance of smaller prey items (Fig.~\ref{fig:PPMR_feedinglevel}.a). The realised mean PPMR calculated from EQT (using ${\cal N} \propto m_p^{-\lambda}$) is $\exp[ (\lambda -3/2)\sigma^2 ] \beta \approx 1.7\beta$. Realised PPMR from the simulations oscillate around this value due to the fluctuations in the community spectrum (Fig.~\ref{fig:spectra}).

As individuals grow to a size larger than $\beta m_0$ they switch from eating food in the resource spectrum to feeding on other species. This leads to a decrease in the feeding level from $f_0 = 0.6$ to about 0.45. The oscillations in feeding level increase in magnitude as body size increases due to larger fluctuations in prey availability (Fig.~\ref{fig:PPMR_feedinglevel}.b). Many large individuals periodically have a feeding level below the critical feeding level $f_c$ (where starvation kicks in) since prey items in the preferred size range become scarce, which results in ingestion of smaller food items and therefore increasing PPMR.

\begin{figure}[tb]
  \centering
  \includegraphics[width=12cm]{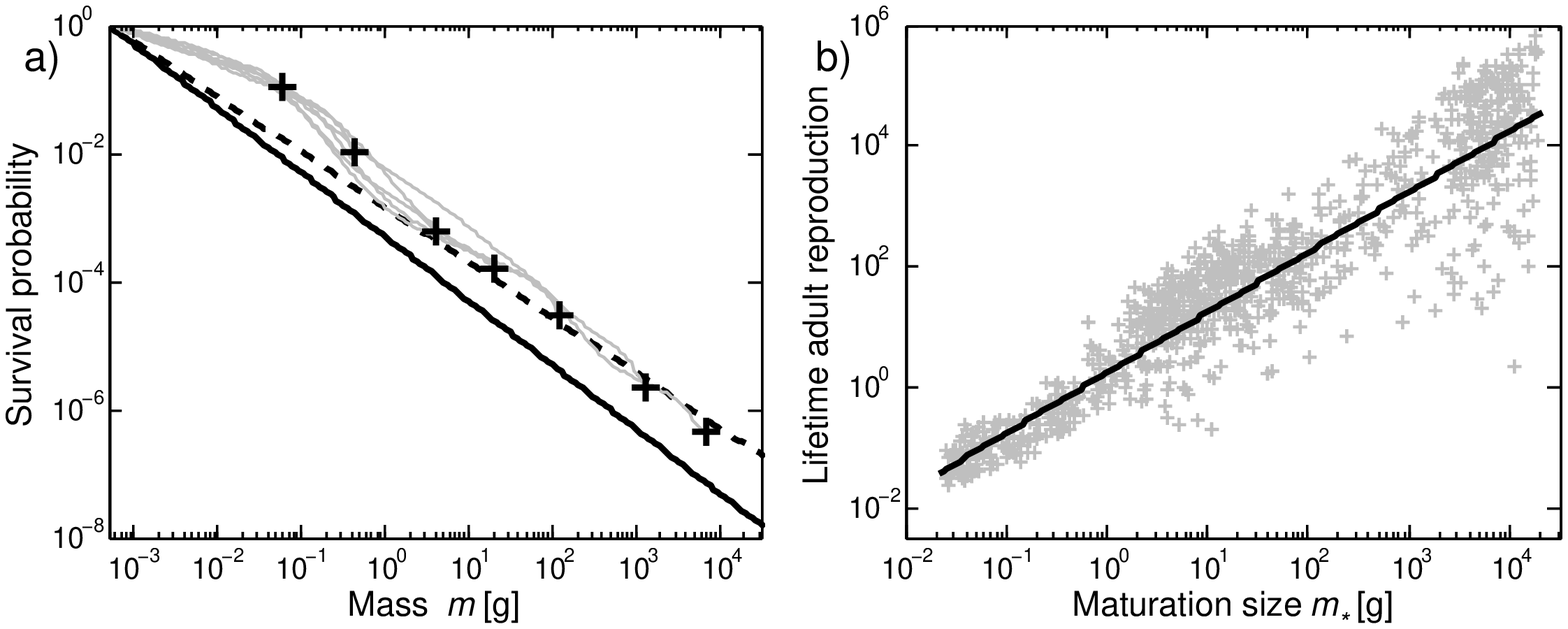}
  \caption{The components of the expected lifetime reproductive output. a) Probability of surviving to maturation size $m_*$ (crosses) along with survival curves throughout life (grey line) for data pooled into 7 logarithmic groups. Expected $p \propto m_*^{-a}$ scaling from EQT is shown for $a=0.86$ (dashed) and $a=1$ (solid). b) Lifetime adult reproduction for all species (crosses) along with EQT $R_\mathrm{adult}\propto m_*$ expectation (line).}
  \label{fig:survivalgraphs}
\end{figure}

\subsection{Reproduction and survival}
Lifetime adult reproduction calculated from the simulated food webs fit the EQT prediction since it scales linearly with $m_*$ (Fig.~\ref{fig:survivalgraphs}.b). The probability of surviving to a given size is independent of $m_*$, as the survival curves of the different $m_*$ groups lie on top of one another (Fig.~\ref{fig:survivalgraphs}.a). Survival to $m_*$ scales inversely with $m_*$  (i.e.~$a=1$, crosses in Fig.~\ref{fig:survivalgraphs}.a), which ensures that $R_0$ is constant. However, if the $a=1$ scaling of survival to $m_*$ is extrapolated to $m_0$ it is seen that it does not intersect $p_{m_0\rightarrow{}m_0}=1$. Instead the survival curves change slope between $m_0$ and around $10^{-1}$\,g where predation mortality starts to dominate due to an abundance of fish individuals in the same order of magnitude as the resource spectrum, which is intensified by reduced growth stemming from food competition (Fig.~\ref{fig:spectra}). In summary survival does not scale with $m_*^{-a}$ as predicted by EQT. Instead adult survival scales with $m_*^{-1}$ (i.e.~$a=1$) whereas individuals smaller than $\approx 0.1\,$g have a higher survival (i.e.~a smaller scaling exponent).

\section{Discussion}
We have developed a generic food web framework suitable for analysing systems of interacting size-structured populations. The framework increases ecological realism compared to traditional unstructured food web models by explicitly resolving the whole life-history of individuals, but maintains simplicity by describing species with only one trait: maturation size $m_*$. Remaining parameters are made species-independent through inter- and intraspecies scaling with $m_*$ and body mass $m$. The productivity of the system is characterised by one parameter, the initial feeding level $f_0$. Feeding behaviour is assumed to be determined by a feeding kernel with a fixed preferred PPMR (big individuals eat small individuals), multiplied by a species-specific coupling strength.

Only characterising the life-history and feeding preference of individuals of a species by body mass $m$ and trait $m_*$ is clearly a simplification, but contemporary knowledge suggests that a large part of the individual bioenergetics related to growth \citep{book:Peters1983} and reproduction \citep{bib:Blueweiss_etal1978} indeed can be described by such scaling. Additionally it is well-known that predators often outsize their prey \citep{bib:Brose_etal2006a} which justifies the use of the generalisation ``big ones eat small ones''.

\subsection{Model architecture} \label{sec:disc_model}
The model was parameterised from cross-species analyses of fish communities, since aquatic systems constitute a group of strongly size-structured ecosystems. Other less strongly size-structured taxa can be modelled as well through reparametrisation and by allowing each species to have its own offspring size $m_{0,i}$. Additionally,  the description of how available energy energy is divided between growth and reproduction may have to be reformulated since animals in other taxa may exhibit determinate growth. Determinate growth can be modelled simply by replacing the allocation function \eqref{eq:psi-function} with only the part within the square brackets.

The proposed modelling framework is similar to physiologically structured models \citep{bib:Andersen_and_Ursin1977,book:Metz_etal1986,bib:de_Roos_and_Persson2001}, and as these based on mechanistic individual-level processes. Our contribution is to employ a trait-based description of species identity, and a formulation of food preference which is split into a size- and a species-based contribution, which renders the developed framework useful as a generic food web framework. Recently the PSPM approach has been reduced to a stage-structured model which facilitates multi-species studies \citep{bib:de_Roos_etal2008}; however this is achieved at the cost of collapsing continuous size-structure to a discrete stage-structure. A first step towards multi-species PSPMs was carried out with an intra-guild predation model, which showed that obtaining species coexistence between two size-structured populations is a difficult task \citep{bib:Wolfshaar_etal2006}; a result which is probably due to insufficient ecological differentiation of the two species. In the proposed framework the trait maturation size provides a simple and logical way of representing ecological differentiation of species, whereas this differentiation in PSPMs is less clear due to large species-specific parameter sets. Additional ecological differentiation and heterogeneity are obtained by also including food web structure in the form of species coupling strengths.

An alternative approach to model a size-structured community is the community size-spectrum models \citep{bib:inbookKerfoot-Silvert_etal1980,bib:Benoit_and_Rochet2004}. In these models the community is represented by a community size-spectrum of all individuals irrespective of species identity \citep{bib:Sheldon_and_Parsons1967}. As with the physiologically structured models these are based on individual-level descriptions of life-history. The community spectrum approach has the drawback that species are not resolved, as all individuals are lumped together into one spectrum. Their advantage is their ability to make community-wide predictions with simple means \citep{bib:Blanchard_etal2009} similarly to the mean-field theory in unstructured food webs \citep{bib:McKane_etal2000,bib:Wilson_etal2003}.

A central element in the model is the division of energy between somatic growth and reproduction through the allocation function $\psi(m,m_*)$. As in PSPMs our bioenergetic model is a net-production model where it is assumed that metabolic costs are covered with highest priority after which the remaining energy can be used for growth and reproduction. PSPMs are formulated either with one state variable: individual body weight \citep{bib:Kooijman_and_Metz1984, bib:Claessen_and_de_Roos2003}, or with two state variables: somatic weight and reserve weight \citep{bib:de_Roos_and_Persson2001}. In the latter case energy is divided between the two states such that the ratio between the two state variables is aimed to be constant, and accumulated reserves are used for reproduction at the end of the growing season. In the case with only one state variable surplus energy is divided between somatic growth and reproduction with a fixed ratio ($\kappa$-rule). When using the $\kappa$-rule the maximum asymptotic size any species individual can obtain is $M_+$ where intake $\alpha{}hf(M_+)M_+^n$ equals the metabolic costs $kM_+^p$ -- meaning that all species would obtain the same asymptotic size if parameters are species independent as in our formulation. $M_+$ is very sensitive to the precise values of $n$ and $p$, and they can therefore only be regarded as poor determinators for asymptotic size \citep{bib:Andersen_etalMP2008}. Our model deviates from the single-state PSPMs in this partitioning of energy, as we assume that mature individuals allocate an amount proportional to their body size for reproduction \citep{bib:Blueweiss_etal1978}, and that asymptotic size depends on the trait size at maturation \citep{bib:Beverton1992,bib:Froese_and_Binohlan2000,bib:He_and_Stewart2001}. This ensures that the ratio between gonad size and somatic weight is constant within a species, which is similar to the partitioning rule used in two-state PSPMs. The allocation function is derived under the assumption of a constant feeding level throughout adult life. Even though the feeding level is assumed constant, the actual allocation still vary depending on the actual food availability, as $\psi(m,m_*)$ only determines the fraction of available energy allocated to reproduction. An alternative way to derive $\psi(m,m_*)$ is to let it depend on actual feeding levels. This assumption, however, would imply that individuals adjust their allocation to reproduction such that asymptotic size is always reached. This does not seem plausible as individuals in resource scarce environments probably obtain smaller maximum sizes, and therefore we find the most reasonable assumption to be that of a constant feeding level. The exponents $n$ and $p$ are close to equal in nature, and for $n=p$ the energy allocation function leads to biphasic growth where adults follow von Bertalanffy growth curves \citep{bib:Lester_etal2004}. We fixed the yearly mass-specific allocation to reproduction (yearly gonado-somatic index, GSI) to be independent of individual body size within a species. The obtained form of $\psi(m,m_*)$, however, yields a $m_*^{n-1}$ scaling of yearly GSI across species, which is consistent with empiric evidence \citep{bib:Gunderson1997}. This means that the form of $\psi(m,m_*)$ implies a trade-off between $m_*$ and the mass-specific reproduction: large $m_*$ species can escape predation mortality via growth by paying the price of a lower mass-specific reproduction \citep{bib:Charnov_etal2001}. When the exponents $n$ and $p$ differ, growth will still be biphasic and adult growth curves will be similar to von Bertalanffy curves \citep[see also][]{bib:Andersen_and_Pedersen2010}. In conclusion the derived allocation rule leads to realistic growth patterns.

\subsection{Food web structure}
Food web structure is the most essential part of a food web model, and in principle two approaches can be taken to obtain a structure for a dynamic food web model: a top-down and a bottom-up approach.

The top-down method generates food web matrices from the desired number of species and connectance using a static model (stochastic phenomenological models: \citet{bib:Cohen_and_Newman1985,bib:Williams_and_Martinez2000,bib:Cattin_etal2004,bib:Allesina_etal2008}, or more mechanistic approaches involving phylogenetic correlations \citep{bib:Rossberg_etal2006} or foraging theory \citep{bib:Petchey_etal2008}). Next, the food web matrix is used to drive a dynamic model, which is simulated forward in time where some of the initial species will go extinct, and the remaining species set can be used for analysis. Note that in addition to a decreased species richness in the final community other food web statistics as e.g.~the final connectance may differ as well \citep{bib:Uchida_and_Drossel2007}.

In the bottom-up approach link strengths are determined from ecological relations, such as e.g.~a predator-prey feeding kernel: if the prey fits into a certain size range relative to the predator size, then interaction occurs between the nodes with a strength determined by the feeding kernel \citep{bib:Loeuille_and_Loreau2005,bib:Virgo_etal2006,bib:Lewis_and_Law2007}. Predator preferences depend, in addition to ecological characters, on evolutionary history and recent approaches add this component of phylogenetic correlations \citep{bib:Rossberg_etal2008}.

For size-structured food webs a top-down algorithm for generating realistic food web matrices does not exist. This is due to lack of data describing the three dimensional interaction matrix -- dimension one and two is respectively predator and prey identity as in the classic interaction matrix, and the third dimension is predator/prey body size. Thus one is confined to the bottom-up approach and/or random interaction matrices. In this study we use the bottom-up approach to prescribe interactions to obey the pattern of ``big ones eat smaller ones''. Life-history omnivory \citep{bib:Pimm_and_Rice1987} is therefore naturally incorporated in size-structured food webs through the use of a feeding kernel. To obtain an ensemble of different communities we use the top-down approach of a classical two-dimensional predator-prey interaction matrix -- that is we assume that regardless of size individuals within a species have equal potential maximum link strength (coupling strengths in our model) to another species. As no top-down method exists for generating this matrix we use random matrices. The actual link strength is the product of the coupling strength and the feeding kernel, meaning that link strengths indeed are dynamic as they depend on the size-structure of both prey and predator.

As we generate food webs from a fixed initial pool of only 30 species and use a random matrix as coupling matrix we only obtain small food webs (maximum: 9 species). However, it should be noted that the number of resource species the resource spectrum represents is not included. To obtain larger food webs a larger species pool is needed along with a sequential assembly algorithm \citep{bib:Post_and_Pimm1983}, and a better method for obtaining coupling strengths between species. Our primary interest in the food web analyses has been the size- and trait-structure of food webs with a finite number of species, and how these compare with EQT predictions, which are based on the premise of a continuum of species. The general correspondence with EQT indicates that the broad-scale patterns are relatively insensitive to how the species-specific coupling strengths (i.e.~food web structure) are specified.  Still, an interesting follow-up study would be focused on the coupling matrix structure, which may more generally be size-dependent, and how the effective food web structure that emerges from the coupling strengths and feeding kernel compares with empiric food webs.

\subsection{Community structure}
We generated an ensemble of size-structured food webs and used averages over these to make general predictions of the structure of fish communities, in particular the size-structure of individual populations, and how these spectra ``stack'' to form the community size-spectrum. In accordance with EQT we find the community spectrum to scale with $\lambda = 2+q-n \approx 2$  meaning that the distribution of biomass as a function of individual body size is close to constant when individuals are sorted into logarithmically evenly sized bins. This prediction means that the biomass of individuals between e.g.~1$\,$g and 10$\,$g is the same as those present with body sizes between 1$\,$kg and 10$\,$kg, in accordance with the Sheldon hypothesis \citep{bib:Sheldon_etal1972}.

The distribution of biomass as a function of $m_*$ is predicted to be almost independent of $m_*$ in accordance with EQT. The result is reminiscent of the Sheldon hypothesis, and it can be formulated as an extension of the Sheldon hypothesis: \emph{``The total biomass of individuals ordered in logarithmically spaced groups of their maturation size is approximately constant''}. This means that the total biomass of all species with $m_*$ between 1 and 10 g is approximately the same as that of species with $m_*$ within 1 to 10 kg. This prediction is a novel extension and could be tested by size-based field data. In contrast to EQT the dynamic framework also provides predictions on the distribution of the number of species as a function of $m_*$. Species tend to cluster in groups on the $m_*$ axis separated by a distance corresponding to the preferred PPMR $\beta$. This clustering is partly a reflection of the use of a fixed value of $\beta$; more diversity in feeding strategies (i.e.~different $\beta$) would probably smoothen the species distribution as well as making the feeding level more constant.

The size-spectra of individual species do not to follow power laws as predicted by EQT since there is a change in spectrum slopes from small to medium sized individuals. This difference stems from different scaling relationships for the survival probability of small and larger individuals. The less steep slope in survival for small individuals is due to a proportionally low mortality rate caused by their low abundance relative to similarly sized resource items. Incorporating mortality from the resource spectrum on the smallest individuals may thus result in a single survival probability scaling. The probability of surviving to $m_*$ scale as $m_*^{-a}$ for a physiological predation constant value of $a=1$, which is conflicting with the value $a=0.86$ predicted by EQT. The discrepancy about the value of $a$ highlights an inconsistency within EQT: Enforcement of mass-balance between growth and predation leads to $a=0.86$, while the reproductive boundary condition can only be fulfilled if $a=1$. The full food web simulations demonstrate that both the scaling of surviving to $m_*$ and the scaling of the number of offspring are best predicted by a value of $a=1$.  This indicates that when EQT predictions depend on $a$, the value $a=1$ should be used even though that breaks mass conservation in EQT.

Lastly we demonstrate that realised PPMRs (i.e.~PPMRs based on ingested prey) emerge in the model. Average realised PPMR is always larger than the preferred PPMR $\beta$ since smaller prey items are more abundant than larger ones. It is found that the realised PPMR is proportional to the preferred ratio ($PPMR=1.7\beta$). Model predictions show that realised PPMR oscillates around this value due to fluctuations in the average community spectrum. PPMR displays large fluctuations with size demonstrating that determination of PPMR from single measurements is problematic due to high prey abundance sensitivity. Empirical findings show that realised PPMRs increase with body size \citep{bib:Barnes_etal2010}, but one should be careful about concluding that the preferred PPMR (which we put into models) shares this size scaling, since relative abundances may cause the increase rather than actual behavioural prey preferences: even though we have a fixed preferred PPMR our model predicts that realised PPMR is an increasing function of body size.

\subsection{Conclusion and outlook}
The proposed food web framework increases ecological realism in food web models as it resolves the complete life-history of individuals by representing the size-structure of each species with a size-spectrum. More specifically the framework complies with five requirements of (cf.~Introduction): 1) being generic with few parameters, 2) being mechanistic and utilising individual-level processes, 3) including food dependent growth, 4) being practically solvable for species-rich communities, and 5) complying with data on community structure and individual growth curves.

Trait-based size-structured food webs can be examined at four levels of organisation: at community level, at species level, at trait level, and at the individual level. We generated empirically testable hypotheses of mainly biomass distributions at different levels of organisation.

By assuming a power law community spectrum and a constant feeding level the full dynamic model can be simplified to an EQT model \citep{bib:Andersen_and_Beyer2006}. Correspondence of predictions by EQT and the full model justifies the use of the simplifying assumptions. EQT is a powerful analytical tool that in a simple manner yields insight to e.g.~the biomass distributions within and across species in size-structured food webs. However, as EQT assumes steady-state, the study of emerging effects, e.g.~diversity and responses to perturbations, has to be conducted with the full model.

The PSPM framework has showed existence of alternative stable states where single populations can exist with different size-structure compositions \citep{bib:de_Roos_and_Persson2002,bib:Persson_etal2007b,bib:de_Roos_etal2008PNAS}. It is an open question whether such alternative states become more widespread or if they disappear when more species interact with each other. This question is important since it tells whether such alternative states are expected to occur frequently or rarely in nature, and consequently whether exploitation can easily induce shifts between states. An important future challenge is thus to study the possibilities of multiple states in complex food webs -- not only of single individual populations, but of the ecosystem as a whole. The proposed framework allows exactly this kind of studies since it provides a full ecologically realistic but conceptually simple model of size-structured ecosystems.

Natural future extensions of the model could be to allow the species coupling strengths to be size-dependent and make coupling strengths depend on vulnerability and forageability of prey and predators \citep{bib:Rossberg_etal2008} as well as on the spatial overlaps of the interacting species. Adding this extra level of mechanistic realism would allow the framework to be useful for studying ecosystem consequences of spatial changes of species populations, which could be driven by climatic changes.

\section*{Acknowledgements}
\begin{small}
Niels Gerner Andersen is thanked for valuable discussions on bioenergetic models. MH was supported by the European Marie Curie Research Training Network FishACE (Fisheries-induced Adaptive Changes in Exploited Stocks) funded through the European Community's Sixth Framework Programme (Contract MRTN-CT-2004-005578). KHA was supported by the EU 7th framework research projects MEECE and FACTS.
\end{small}


\ifthenelse{\boolean{BIBUNITRUN}}{ \begin{small} \putbib \end{small} \end{bibunit} }{ \begin{small} 
 \end{small} }

    \newpage
    \processdelayedfloats 








    \cleardoublepage
    \ifthenelse{\boolean{BIBUNITRUN}}{ \begin{bibunit} }{ }
    \setcounter{page}{1}
    \appendix
    \begin{small}
    {\Large \emph{Appendix should be available as online material.}}

\section{Derivation of predation mortality} \label{app:pred-mortality}
Predators with a size between $m$ and $m+dm$ have a food intake rate of $s(m_p,m) f(m) hm^n \theta N(m)dm$ for $m_p$ sized prey. The total density of food available from all prey sizes to the predators in $[m \, ; m+dm]$ is $\phi(m)$ \eqref{eq:available_food}, meaning that the mortality experienced by a $m_p$ sized individual is:
\begin{equation}
  \mu_{p,i}(m_p) = \sum_j \int \frac{ s(m_p,m) f_j(m) hm^n \theta_{j,i} N_j(m) }{ \phi_j(m) }\, dm.
\end{equation}

The maximum food intake may be expressed as a function of $f(m)$, $v(m)$, and $\phi(m)$ via \eqref{eq:feeding_level}, such that the predation mortality can be written as:
\begin{equation}
  \mu_{p,i}(m_p) = \sum_j \int s(m_p,m) (1-f_j(m)) v(m) \theta_{j,i} N_j(m)\, dm.
\end{equation}

By using the EQT assumptions of constant feeding level and a power law community spectrum (cf. section \ref{sec:methods}) the mortality reduces to $\mu_{p}(m_p) =  \overline{\theta} (1-\overline{f}) \int s(m_p,m) v(m) \kappa_c m^{-\lambda}\, dm$, which can be solved analytically:
\begin{equation}
  \mu_{p}(m_p) = \alpha_p m_p^{n-1}, \label{eq:app:community_mortality}
\end{equation}
\noindent{}where $\alpha_p = \overline{\theta} (1-\overline{f}) \sqrt{2\pi}\kappa_c\gamma\sigma\beta^{1+q-\lambda} \exp\left[\frac{1}{2}\sigma^2(1+q-\lambda)^2\right]$.

\section{Available food and the physiological level of predation $a$} \label{app:avail_food_and_a}
Using the EQT assumption of a power law community spectrum allows calculation of the available food density $\phi(m) = \overline{\theta} \int s(m_p,m) \kappa_c m_p^{-\lambda} m_p\,dm_p$:
\begin{equation}
\phi(m) = \alpha_\phi \overline{\theta}\kappa_c m^{2-\lambda}, \label{eq:app:phi}
\end{equation}
\noindent{}where $\alpha_\phi=\sqrt{2\pi}\sigma\beta^{\lambda-2} \exp\left[\frac{1}{2}\sigma^2(2-\lambda)^2\right]$.

Using the EQT assumption of constant feeding level yielding $\lambda = 2+q-n$ allows us to write $\overline{\theta}\kappa_c=\overline{f}h/(\alpha_\phi\gamma(1-\overline{f}))$ by rearranging the expression of the feeding level \eqref{eq:feeding_level}. Using this and the definition of $\hbar$ allows writing $\alpha_p=c(\hbar+k)\beta^{2n-q-1}/\alpha$ where $c=\exp\left[\frac{1}{2}\sigma^2 \left((1+q-\lambda)^2 - (2-\lambda)^2\right)\right] = 1.03 \approx 1$. Ultimately using the definition of $f_c$ allows writing the physiological level of predation $a=\alpha_p/\hbar$ as:
\begin{equation}
a = c \frac{\overline{f}}{\overline{f}-f_c} \beta^{2n-q-1} / \alpha.
\end{equation}

\section{Calculating efficiency $\epsilon$ of offspring production} \label{app:epsilon_deriv}
The efficiency of turning energy into offspring is denoted $\epsilon$. It includes losses due to behavioural aspects, pre-hatching mortality, and that the energy contents in gonadic tissue is higher than in somatic tissue. It is a quantity that is difficult to measure, but for $n=p$ its value can be derived.

The energy (in units of mass) routed into reproduction (for $n=p$) is  $\psi(m,m_*) \hbar m^{n}$ where $\hbar = \alpha \overline{f} h-k$. The energy of the produced offspring is then, $E_o(m)=\epsilon\psi(m,m_*) \hbar m^{n}$:
\begin{equation}
E_o(m) = \epsilon \hbar \eta_*^{1-n} m_*^{n-1} m .
\end{equation}

\noindent{}From \cite{bib:Gunderson1997} we have the yearly mass-specific allocation to reproduction:
\begin{equation}
\varrho(m_*) = \varrho_0 \eta_*^{1-n} m_*^{n-1}, \label{eq:GSI-Gunderson}
\end{equation}

\noindent{}where $\varrho_0 = 1.2\,\mathrm{g}^{1-n}/\mathrm{year}$ is obtained using least sum of squares in fitting the curve to the data for oviparous fish in \cite{bib:Gunderson1997}. Equalling \eqref{eq:GSI-Gunderson} and $E_o/m$ allow us to determine the efficiency of offspring production $\epsilon$:
\begin{equation}
\epsilon = \frac{\varrho_0}{\hbar} \approx 0.12.
\end{equation}

\section{Setting the search rate prefactor $\gamma$ from initial feeding level $f_0$} \label{app:gamma}
Food for the smallest individuals in the spectra will be supplied by the background spectrum. If we assume that the resource spectrum is at carrying capacity $\kappa$ then an equilibrium initial feeding level $f_0$ for the small individuals can be calculated using \eqref{eq:feeding_level}.

Alternatively we may specify an initial feeding level $f_0$ and derive one other parameter. By solving the feeding level for $\gamma$ by using the analytical solution for the density of food $\phi(m)$ \eqref{eq:app:phi} we find $\gamma$ as a function of $f_0$:
\begin{equation}
\gamma = \frac{f_0 h}{(1-f_0)\alpha_\phi\theta_{i,R}\kappa} \approx \frac{f_0 h \beta^{2-\lambda}}{(1-f_0)\sqrt{2\pi}\sigma\theta_{i,R}\kappa}. \label{app:eq:gamma}
\end{equation}

\section{Parameter estimation} \label{app:parameter_estimation}
\emph{Individual growth:} From \cite{bib:Kitchell_and_Stewart1977} we obtain an estimate of specific dynamic action on 15 \% of food consumption, and conservative estimates of egestion and excretion on 15 \% and 10 \% respectively. This results in an assimilation efficiency of $\alpha=0.6$.

The maximum intake scales with a 0.6--0.8 exponent, with $n=0.75$ being an approximate average value \citep{book:Jobling1994}. \cite{bib:AndersenNG_and_Riis-Vestergaard2004} provides a length-based relationship for the maximum intake rate based on a whiting study adopted for saithe. Using $m=0.01 l^3$ ($m$ in g and $l$ in cm) \citep{book:Peters1983}, and an energy content of 5.8\,kJ/g (fish) or 4.2\,kJ/g (invertebrates) \citep{bib:Boudreau_and_Dickie1992} yields a prefactor $h$ for the maximal food intake on 83\,g$^{1-n}$/year or 114\,g$^{1-n}$/year (at $10^\circ$C). These intake values overestimate the intake of large individuals since \cite{bib:AndersenNG_and_Riis-Vestergaard2004} use an intake exponent of 0.67 instead of $n=0.75$. Due to this a value of $h=85$\,g$^{1-n}$/year is selected, which also provides reasonable fits to 'cod-like' growth curves ($m_*=5$ kg).

The standard metabolism scaling exponent $p$ for fish is slightly higher than for other taxa, around 0.8 from intra- and interspecies  measurements \citep{bib:Winberg1956,bib:Killen_etal2007}. For simplicity we assume $p=n$. The first term (acquired energy) in the growth model \eqref{eq:growth-function} is $\alpha{}f(m)hm^n$ where the feeding level $f(m)$ is a decreasing function of body size (see \textit{Results}). This has the effect that even when $n=p$ is assumed the acquired energy term still effectively scale with a smaller exponent than the maintenance term $km^p$ in accordance with the experimental data on food intake and standard metabolism. Furthermore it is noted that this clearly makes the individuals in each functional species non-neutral. The bioenergetic consequences of $n \neq{} p$ has been explored in detail by \cite{bib:Andersen_etalMP2008}.

The prefactor for standard metabolism can from \cite{book:Peters1983} be determined to $6.5$\,g$^{1-n}$/year if the diet is composed of only invertebrates and $4.7$\,g$^{1-n}$/year if all the energy is from fish. Both values were corrected to $10^\circ$C using $Q_{10}=1.83$ \citep{bib:Clarke_and_Johnston1999}. It is assumed that energy costs due to activity can be described with an activity multiplier on the standard metabolic rate. Estimations of activity costs are difficult to obtain, but activity multipliers are often reported in the range 1 to 3; e.g.~1.25 for North Sea cod \citep{bib:Hansson_etal1996}, 1.7 for dace \citep{bib:Trudel_and_Boisclair1996}, and 1.44-3.27 for saithe \citep{bib:AndersenNG_and_Riis-Vestergaard2004} (however see also \cite{bib:Rowan_and_Rasmussen1996,bib:He_and_Stewart1997}). A reasonable value of the prefactor for the standard metabolism and activity costs is assumed to be $k=10$\,g$^{1-n}$/year corresponding to an activity multiplier in the range 1.5 to 2.1.

\emph{Food encounter:} The preferred predator-prey mass ratio is set to $\beta = 100$ \citep{bib:Jennings_etal2002b} and the width of the selection function to $\sigma=1$, which catches the general picture for at least cod and dab \citep{bib:Ursin1973}. It should be noted that small organisms such as copepods have a larger $\sigma$ of 3--4.5 \citep{bib:Ursin1974}, but for simplicity and since focus is on species with rather large $m_*$ the width $\sigma$ will be held constant.

The exponent for swimming speed is $q=0.8$ \citep{bib:Andersen_and_Beyer2006}. The prefactor $\gamma$ for the volumetric search rate is difficult to assess from the literature. An alternative approach is to determine it as a function of of initial feeding level $f_0$ via \eqref{app:eq:gamma}. Experience with the model shows that feeding level is a decreasing function of body size. This means that it is sensible to use an initial feeding level $f_0$ that is larger than the expected average feeding level $\overline{f}$. In this study a default value of $f_0=0.6$ is used. This along with default parameters yields $\gamma=0.8\cdot10^4\,\mathrm{m^3g}^{-q}$/year (Table \ref{tab:parameters}). An alternative estimate of $\gamma$ is possible by multiplying the prefactors for swimming speed \citep{bib:Ware1978} and swept reactive field area (reactive radius assumed equal to body length): $\gamma = 20.3\cdot{}\pi\cdot0.01^{-2/3}\mathrm{cm^3g}^{-q}/\mathrm{s} \approx 4.3 \cdot 10^4\,\mathrm{m^3g}^{-q}$/year, which indeed justifies the use of $f_0=0.6$.

\emph{Mortality:} Realistic energy reserve sizes may be $\xi \in [5\%;\ 20\%]$, and in the present study $\xi=0.1$ is used. A widely used background mortality for 'cod-like' $m_*=5$\,kg fishes is $\mu_b=0.1$\,year$^{-1}$, which yields $\mu_0=0.84$\,g$^{1-n}/$year.

\emph{Reproduction:} The efficiency of offspring production was not found in the literature. However, an analytical expression may be obtained (for $n=p$) by combining the calculation of yearly mass-specific allocation to reproduction from the bioenergetic model (\ref{app:epsilon_deriv}) with empirical measurements \citep{bib:Gunderson1997}, which yields $\epsilon=\varrho_0/\hbar\approx0.1$.  The fraction of asymptotic size to mature at is $\eta_*=0.25$ \citep{bib:Andersen_etalMP2008}. Offspring mass is $m_0=0.5$ mg which corresponds to an egg diameter of 1 mm \citep{bib:Wootton_rj1979,bib:inbookChambers-Chambers_and_Trippel1997}.

\emph{Resource spectrum:} The carrying capacity of the resource spectrum should equal the magnitude of the community spectrum: $\kappa m^{-\lambda}$, with an exponent $\lambda = 2- n + q = 2.05$ \citep{bib:Andersen_and_Beyer2006}. The magnitude of the resource spectrum is set to  $\kappa = 5 \cdot 10^{-3}$ $\mathrm{g}^{\lambda-1}/\mathrm{m}^3$, which is comparable with findings from empirical studies \citep{bib:Rodriguez_and_Mullin1986}. The constant for resource regeneration rate is $r_0=4$ $\mathrm{g}^{1-p}/\mathrm{year}$ at $10^\circ$C \citep{bib:Savage_etal2004}. The cut-off of the resource spectrum is set to include mesoplankton, $m_{cut}=0.5$ g.

\section{Expected Lifetime Reproductive Success} \label{app:EQT}
The expected lifetime reproductive success can be split into two components: 1) the probability of surviving to become adult, and 2) lifetime reproduction per adult.

\subsection{Survival probability}
If we set $\frac{\partial N}{\partial t}=0$ in \eqref{eq:McKendrick-vonFoerster} we may obtain the steady-state solution as:
\begin{equation}
N(m) = \frac{K(m_*)}{g(m, m_*)} \exp\left(-\int \frac{\mu(m)}{g(m, m_*)}dm\right), \label{eq:app:N_solution}
\end{equation}
\noindent{}where $K(m_*)$ is the constant from the integration along $m$. We notice that the probability of surviving from size $m_0$ to size $m$ is $p_{m_0\rightarrow{}m}=\exp\left(-\int_{m_0}^m \frac{\mu(m')}{g(m', m_*)}dm'\right)$, which along with $p_{m_0\rightarrow{}m_0}=1$ allow us to write the survival probability as:
\begin{equation}
p_{m_0\rightarrow{}m} = \frac{N(m)g(m, m_*)}{N(m_0)g(m_0, m_*)}. \label{eq:app:survivalFromSpectra}
\end{equation}

\subsection{Lifetime adult reproduction}
The amount of energy an adult belonging to a $m_*$ population will spend on reproduction throughout its life is:
\begin{equation}
R_{life}(m_*)=\int_{t_*}^\infty p_{t_*\rightarrow{}t} \psi(m,m_*)E(m) dt, \nonumber
\end{equation}
\noindent{}where $t_*$ is maturation age, and $\psi(m,m_*)E_i(m)$ the rate at which energy is allocated to reproduction. Noting that $g(m,m_*)=\frac{dm}{dt}$ allows us to write this as:
\begin{equation}
R_{life}(m_*)=\int_{m_*}^M p_{m_*\rightarrow{}m} \frac{\psi(m,m_*)E(m)}{g(m,m_*)} dm. \label{eq:app:lifeAdultRepro}
\end{equation}

\section{Details of Numerical Methods} \label{app:numerical_method}
The model is in the form of a series of coupled partial-integro-differential equations \eqref{eq:McKendrick-vonFoerster}, one for each species with the size preference function \eqref{eq:size_selection} being the integral kernel. The equations are of the first order in mass (i.e.~hyperbolic) in which case shocks could be formed in the solutions. However the integral kernel smooths out any discontinuities and the equations can be solved effectively and accurately using a standard semi-implicit upwind finite-difference scheme for solving PDEs \citep{book:Num-recipes}. The McKendrick-von Foerster PDE \eqref{eq:McKendrick-vonFoerster} is discretised by calculating $g(m,m_*)$ and $\mu(m)$ explicitly and making the time update implicit, to yield:
\begin{equation}
\frac{N_w^{i+1} - N_w^i}{\Delta{}t} + \frac{g_w^i N_w^{i+1} - g_{w-1}^i N_{w-1}^{i+1}}{\Delta{}m_w} = -\mu_w^i N_w^{i+1} , \label{eq:McKendrick-vonFoerster-discrete}
\end{equation}

\noindent{}where $i$ denotes the time step, and $w$ the grid index on the mass axis ($i,w \in \aleph^+$). First order approximations have been used for both the time and mass derivatives. The discretisation in mass is known as the upwind approximation since the derivative is calculated from $w$ and $w-1$, which is possible since the growth function is non-negative. It is further noted that the $\partial m$ approximation is semi-implicit since densities at time step $i+1$ are used. Equation \eqref{eq:McKendrick-vonFoerster-discrete} may be written as:
\begin{equation}
N_{w-1}^{i+1} \underbrace{\left( -\frac{\Delta{}t}{\Delta{}w_w}g_{w-1}^i\right)}_{A_w} + N_w^{i+1} \underbrace{\left(1 + \frac{\Delta{}t}{\Delta{}w_w} g_w^i + \Delta{}t \mu_w^i \right)}_{B_w} = \underbrace{N_w^i}_{C_w} ,
\end{equation}

\noindent{}which allows us to write an explicit solution for the density spectrum at the $i+1$ time step:
\begin{equation}
N_w^{i+1} = \frac{C_w - A_w N_{w-1}^{i+1}}{B_w},
\end{equation}
which can be solved iteratively since $N_1^{i+1}$ is given by the boundary condition. The flux in the boundary $g(m_0,m_*)N(m_0,t)$ is equal to the reproduction flux of new recruits $R$ \eqref{eq:BC} such that $g_{0}^i N_{0}^{i+1}=R$, which yields: $A_1=0$, and $C_1=N_1^i + \frac{\Delta{}t}{\Delta{}m_1}R$.

This semi-implicit upwind scheme is very stable but diffusive. The third order QUICK (Quadratic Upwind Interpolation for Convective Kinematics) scheme along with the techniques by \cite{bib:Zijlema1996}, which prevents overshooting problems introduced by the QUICK method, were used to check that numerical diffusion poses no problem for $\Delta{}t=0.02$ years, and a $m_w$ mass grid with 200 logarithmically evenly distributed points. To ensure stability the Courant condition (i.e.~\cite{book:Num-recipes}):
\begin{equation}
\frac{|g_w^i|\Delta{}t}{\Delta{}m_w} \leq 1 , \label{eq:courant-condition}
\end{equation}

\noindent{}is prudent to fulfill. The essence of the criterion is that $\Delta{}t$ should be small enough not to allow individuals to skip any mass cells $m_w$ during their growth trajectory.

The grid $m_w$ spans the offspring size $m_0$ to 85\,kg to include maturation sizes up to the order of 20\,kg. The grid for the background spectrum ends at $m_{cut}$, and the lower limit should be low enough to ensure food items for the smallest individuals in the functional species, i.e.~3 decades lower than $m_0$. Identical $\Delta{}m_w$ is used for the background and species spectra to ease computations in the overlap $[m_0;\,m_{cut}]$.

To save computational time the ODEs for the background spectrum \eqref{eq:nb} are solved analytically. The solution at time $t_0+\Delta{}t$ for the semi-chemostatic equation is:
\begin{equation}
N_R(m,t_0+\Delta{}t) = K(m) - \Big(K(m) - N_R(m,t_0)\Big)e^{-[r_0m^{p-1}+\mu_p(m)]\Delta{}t},
\end{equation}

\noindent{}where $K(m) = \frac{r_0m^{p-1}\kappa m^{-\lambda}}{r_0m^{p-1}+\mu_p(m)}$ is the effective carrying capacity at resource size $m$.


    \end{small}
    \ifthenelse{\boolean{BIBUNITRUN}}{ \begin{small} \putbib  \end{small} \end{bibunit} }{ \begin{small} 
 \end{small} }

\end{document}